# Q-ESP: a QoS-compliant Security Protocol to enrich IPSec Framework

Mahmoud MOSTAFA, Anas ABOU EL KALAM, Christian FRABOUL
Université de Toulouse, INPT-ENSEEIHT, IRIT-CNRS, 2 rue Camichel Toulouse –France.
{firstname.lastname}@enseeiht.fr

*Abstract*—IPSec is a protocol that allows to make secure connections between branch offices and allows secure VPN accesses. However, the efforts to improve IPSec are still under way; one aspect of this improvement is to take Quality of Service (QoS) requirements into account. QoS is the ability of the network to provide a service at an assured service level while optimizing the global usage of network resources. The QoS level that a flow receives depends on a six-bit identifier in the IP header; the so-called Differentiated Services code point (DSCP). Basically, Multi-Field classifiers classify a packet by inspecting IP/TCP headers, to decide how the packet should be processed. The current IPSec standard does hardly offer any guidance to do this, because the existing IPSec ESP security protocol hides much of this information in its encrypted payloads, preventing network control devices such as routers and switches from utilizing this information in performing classification appropriately. To solve this problem, we propose a QoS-friendly Encapsulated Security Payload (Q-ESP) as a new IPSec security protocol that provides both security and QoS supports. We also present our NetBSD kernel-based implementation as well as our evaluation results of Q-ESP.

## I. INTRODUCTION

In today's business environments, users demand seamless connectivity and stable access to servers and networks wherever they are: hotels, airports, remote offices, etc. While these functionalities are useful for business, they work only if we can minimize the security risks of transferring sensitive data across the Internet. In order to achieve this goal, there are various security mechanisms for network environment; the most popular is Security Architecture for IP (IPSec) [1]. IPSec is designed to provide inter-operable cryptographically-based security for lPv4 and lPv6. IPSec operates at the IP layer, making it transparent to applications and users.

Unfortunately, security does not come for free and, in general, protection mechanisms require more processing time and causes traffic delay. Real-time applications such as video conferencing, VoIP, and real-time video need special processing to achieve their goals and to overcome the delay introduced by adding security mechanisms. Quality of service (QoS) has been emerged to solve a part of this problem by providing priority treatment to real time traffic. In the QoS domain, the Class of Service concept divides the network traffic into different classes and provides a class-dependent service to each packet. To classify packets, each packet is assigned a priority value. The latter is stored in the "Type of Service" (ToS) field in the IP v4 header (also called Traffic Class in IP v6).

However, allowing the sending device to classify traffic or to set traffic priorities may be subject to threats, as the sender may classify his traffic in a way that gives him upper priorities. This is clearly the disadvantage of what is called *passive admission control*. Conversely, service providers perform *active admission control* by allowing edge routers (neither users nor the sending devices) to inspect the incoming traffic and classify it. The component in charge of this task is called *Multi-Field classifier* (*MFC*) [2].

The inspected fields belong to different network layer headers [3]:

- at Transport Layer Protocol Header: in order to identify the applications running over TCP or UDP, the source and destination port numbers are inspected.
- at Network Layer Protocol Header: *MFC* inspects the source host IP address, the destination host IP address and the protocol identifier (that is used to identify the transport-layer protocol in use).

The previously mentioned five fields of the transport and network protocols headers are used to define the traffic flow [4]. They are collectively called "*five-tuple*". Unfortunately, even if these fields are required for QoS processing, some of them are hidden (encrypted) when using IPSec ESP [5] security protocol. IPSec ESP protocol encrypts the transport layer header, and thus hides the source and destination port numbers as well as the protocol identifier.

To solve this problem, we introduce the Q-ESP "*QoS friendly Encapsulated Security Payload*" protocol that enforces both QoS and security requirements. The major aim of our Q-ESP is to construct packets that are QoS controllable according to active admission control while providing the same security services ensured by IPSec ESP and AH [6].

The remainder of this paper is organized as follows: Section II presents our Q-ESP protocol packet structure. Then, Section III gives the details of Q-ESP processing. Section IV focuses on Q-ESP implementation. Afterward, Section V, provides our Q-ESP analytical evaluation. In Section VI, we present the performances evaluation experiments and results. Finally,

Section VII draws up conclusions and future works.

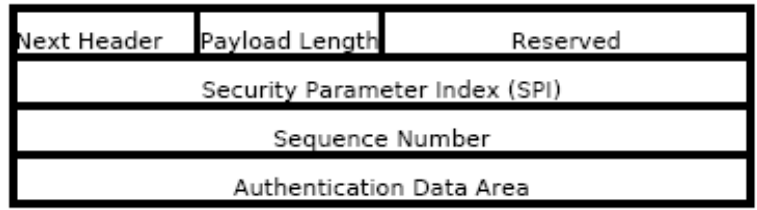

Fig. 1. The AH format.

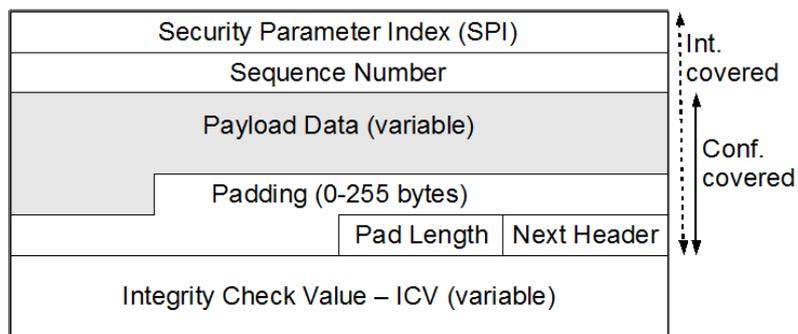

Fig. 1. The ESP format.

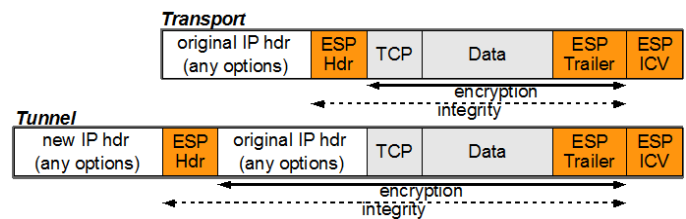

Fig. 3. ESP in transport mode and tunnel mode.

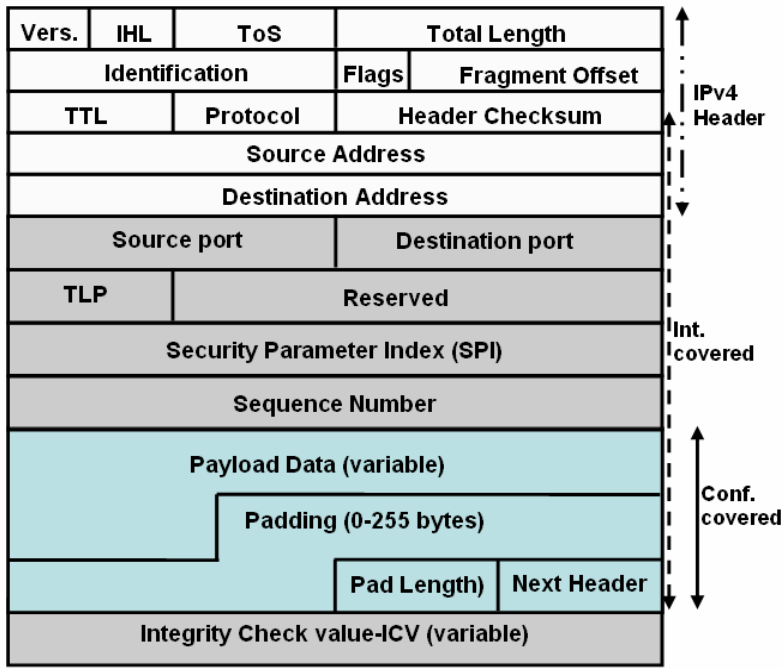

Fig. 4. Q-ESP packet format in Ipv4.

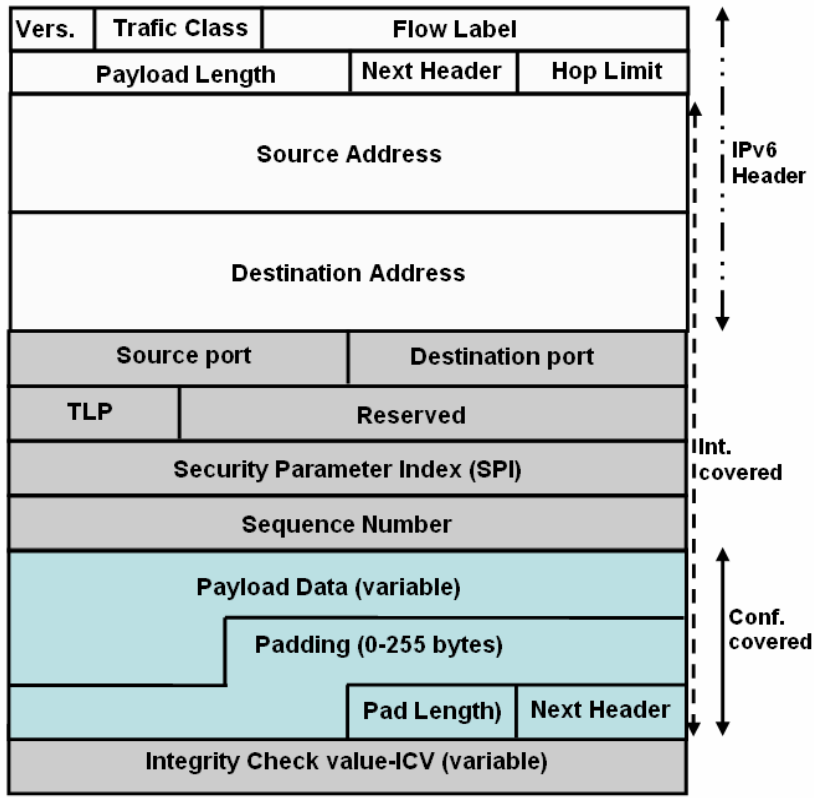

Fig. 5. Q-ESP packet format in Ipv6.

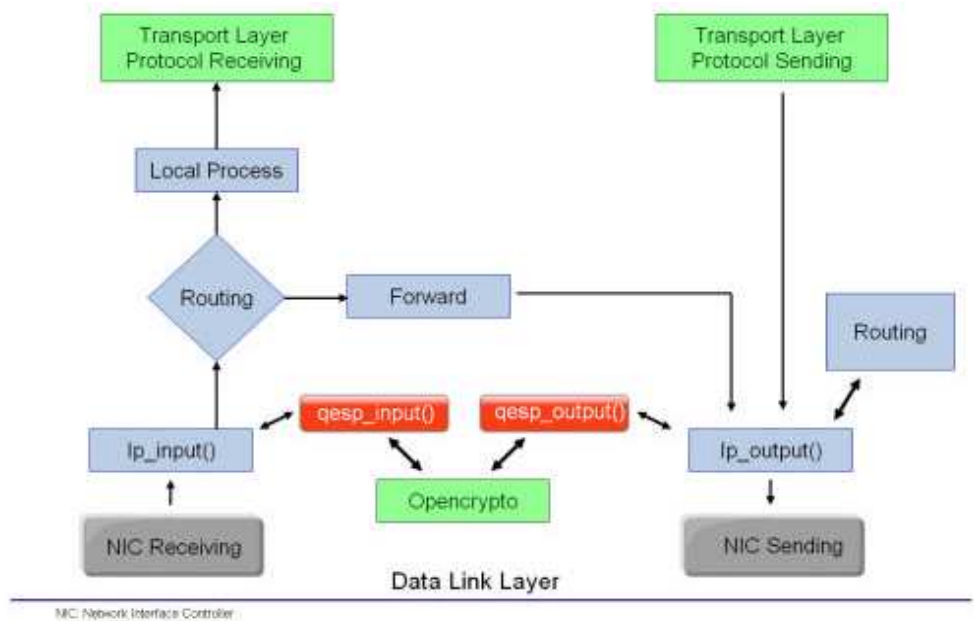

Figure 6. Q-ESP implementation.

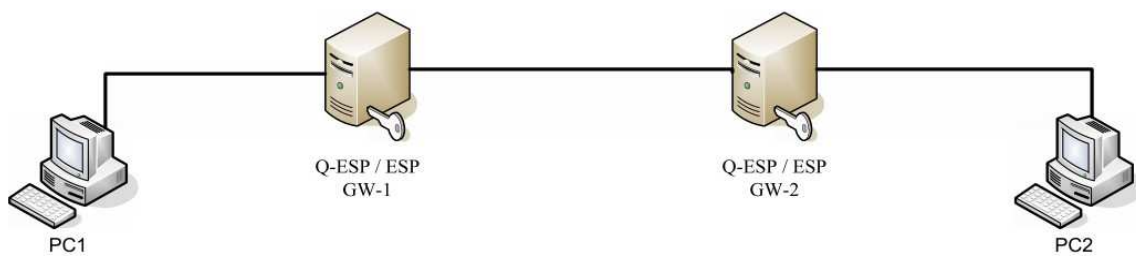

Fig. 7. Testbed for throughput experiments.

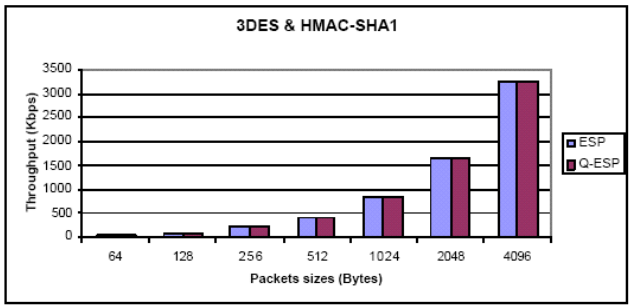
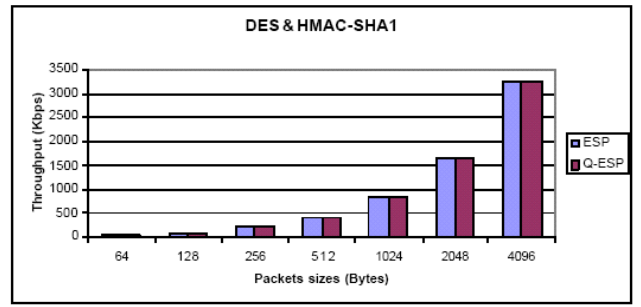
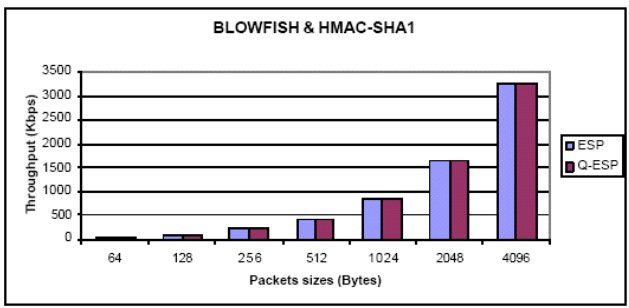
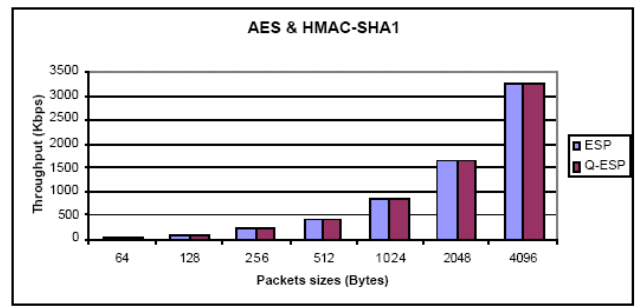

**Figure 8. Throughput using different (encryption/authentication) algorithms for ESP and Q-ESP**

TABLE I: Q-ESP AND ESP THROUGHPUT (KBPS).

| Packet size | ESP | Q-ESP |
|---|---|---|
| 64 | 51.243 | 51.191 |
| 128 | 102.366 | 102.366 |
| 256 | 204.715 | 204.834 |
| 512 | 409.600 | 409.463 |
| 1024 | 819.268 | 818.654 |
| 2048 | 1638.127 | 1637.444 |
| 4096 | 3275.435 | 3275.162 |

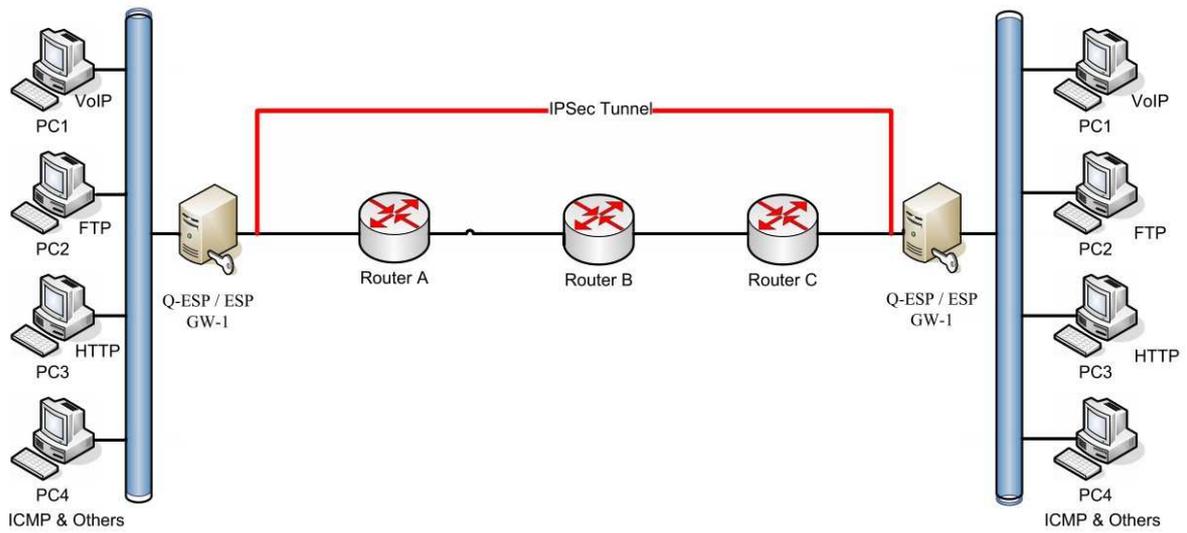

Fig. 8. Testbed for priority control experiments.

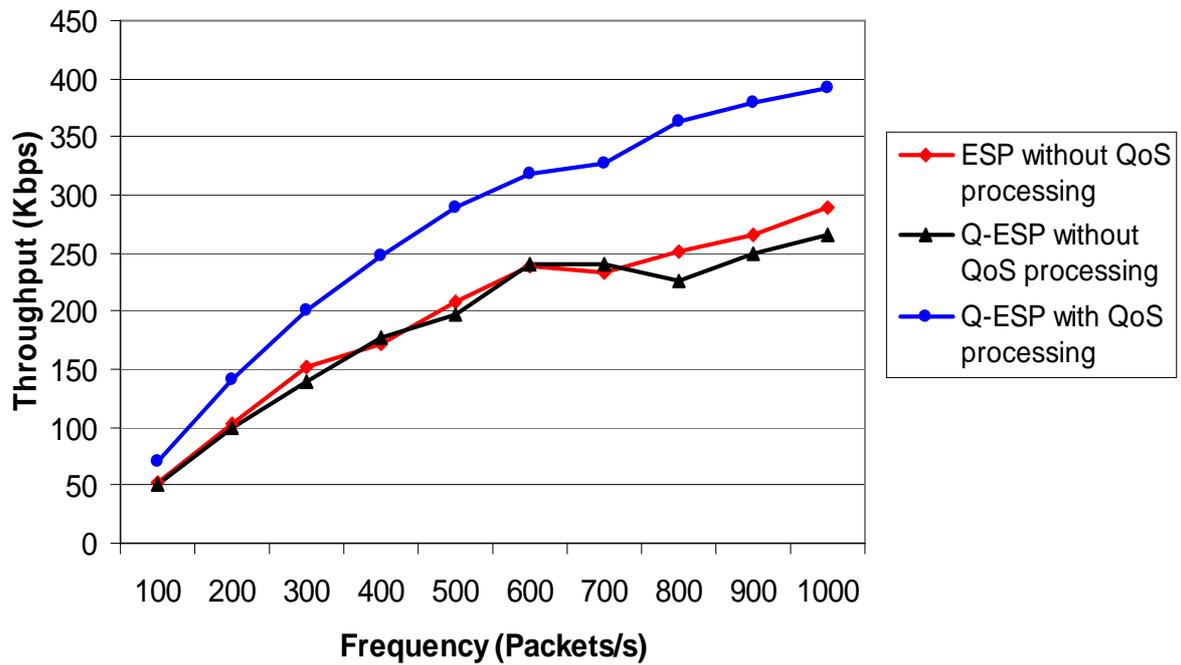

Fig. 9. Priority control experiment results